\documentclass[10pt,a4paper]{article}
\title{Soft colour interactions in hadron-hadron hard diffraction}
\author{A.V. DMITRIEV, N.V.RADCHENKO \\
Novgorod State University, B. S.-Peterburgskaya Street 41,\\
Novgorod the Great, Russia,\\ 117259}
\usepackage{graphicx}
\begin{document}
\maketitle

\abstract{ SCI model gives a good and natural description of DDIS
cross-sections. Howevere, this model is pure phenomenological, and
does not explain the nature of soft color rearrangment. In this
paper we argue, that the most capabilities of SCI model can be
derived from low constituent model and overview applicability of
low constituent model to DDIS processes. }

\section*{}

Soft colour interaction (SCI) model \cite{Edin:1995gi} naturally
explains how rapidity gaps are formed in $ep$ collisions. Pomeron
model is based on another, more phenomenological, approaches.
Moreover, one needs additional contractions (survival gap
probability factor) to fit CDF data at pomeron model.

The ratio of diffractive to non-diffractive events was extracted from SCI model and shows a good agreement with data. More complex observables can not be described without more rigourus theory.

One can see, that SCI model is simplified approximation of LCM  three-stage model of hadron interaction at the high energies.

On the first stage before the collision there is a small number of
partons in hadrons. Their number, basically, coincides with number
of valent quarks and slow increases with the rise of energy due to
the appearance of the  bremsstrahlung gluons.

On the second stage the hadrons interaction is carried out by
gluon exchange between the valent quarks and the initial
(bremsstrahlung) gluons and the hadrons gain the colour charge.
This gluon exchange is time-short Coulomb-like. This soft gluon is
the origin of SCI process, but modern generalized area low (GAL)
version of SCI is not compatible with this description. Main
difference is that SCI model characterize describe soft process
simply by probability of soft rescattering, while we argue that
soft recattering is one-gluon exchange with modified propagator.

On the third stage after the interaction the colour hadrons fly
away and when the distance between them becomes more than the
confinement radius $r_c$, the lines of the colour electric field
gather into the tube of the radius $r_c$. This tube breaks out
into the secondary hadrons.

Because the process of the secondary hadrons production from
colour tube goes with the probability 1, module squared of the
inelastic amplitudes corresponds to the elastic amplitude with the
different parton numbers.

\begin{figure}
\includegraphics[width=2.5in]{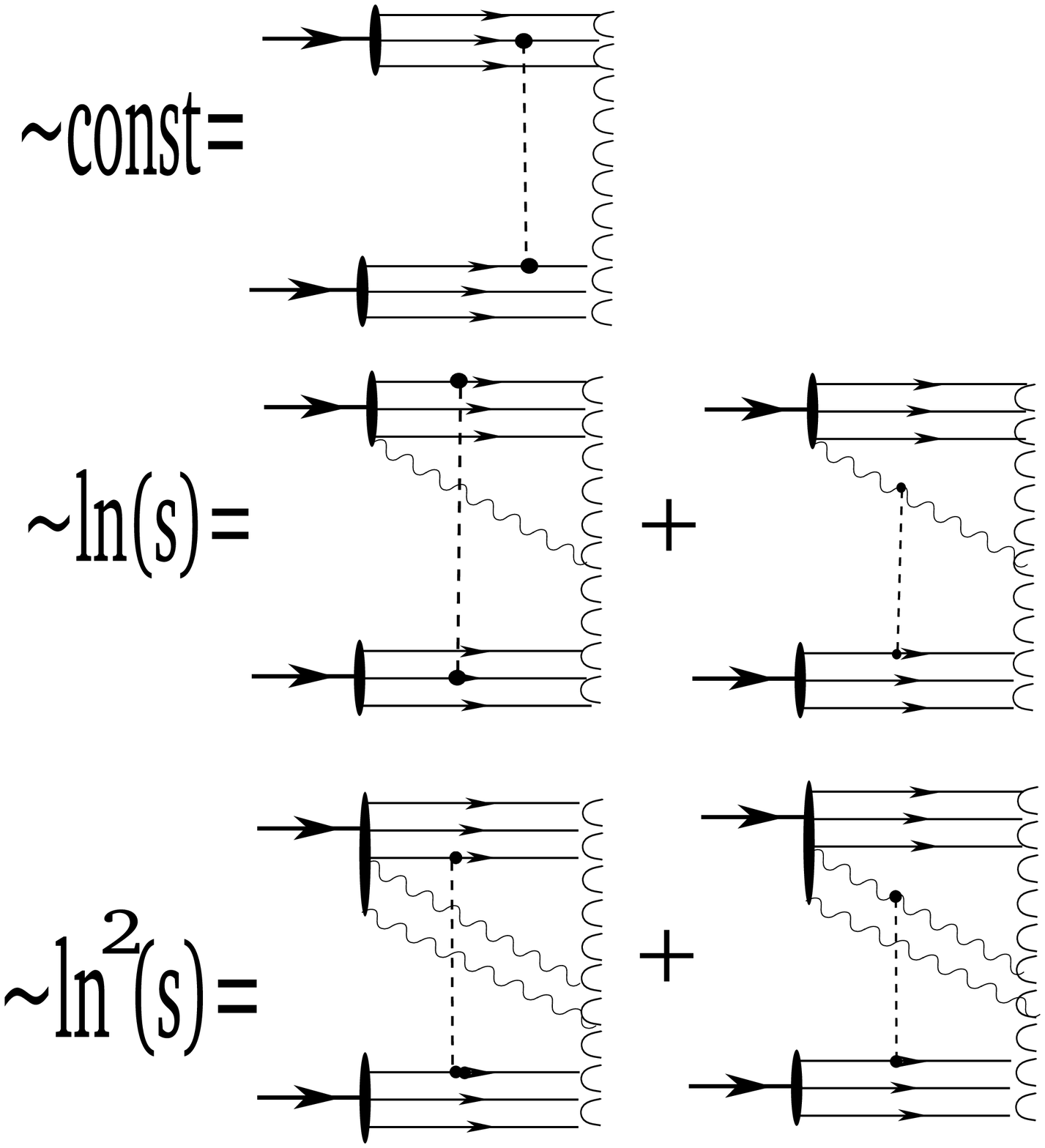}
\vline
\hspace{0.1in}
\includegraphics[width=2.5in]{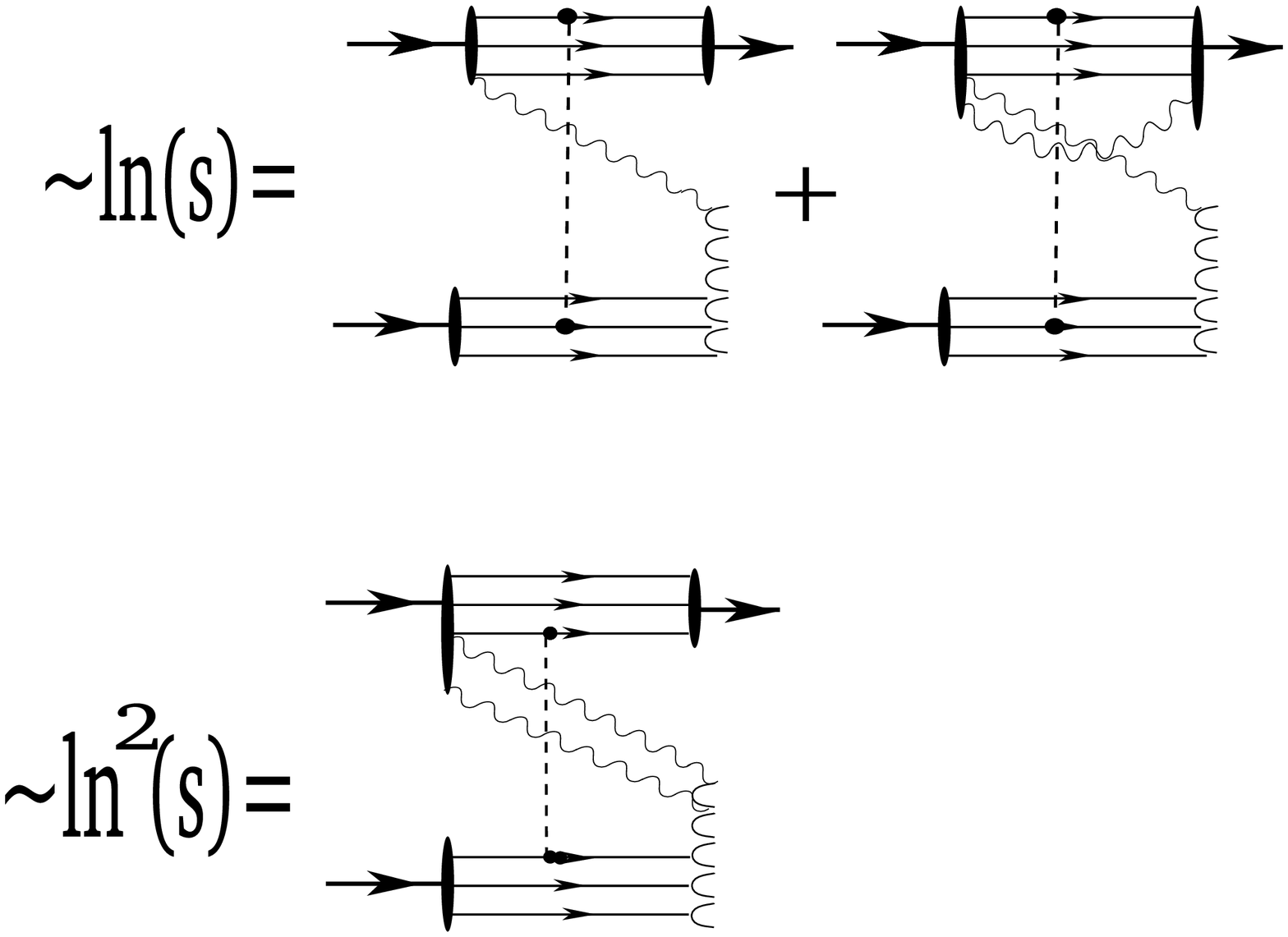}
\caption{Leading by powers of $ln(s)$ diagrams in low constituent model for total cross section (left part) and single diffraction cross section (right part).}
\label{fig:sn}
\end{figure}

This processes are schematically drawn on the left side of Fig.\ref{fig:sn}.

\begin{figure}
\begin{center}
\includegraphics[scale=0.8]{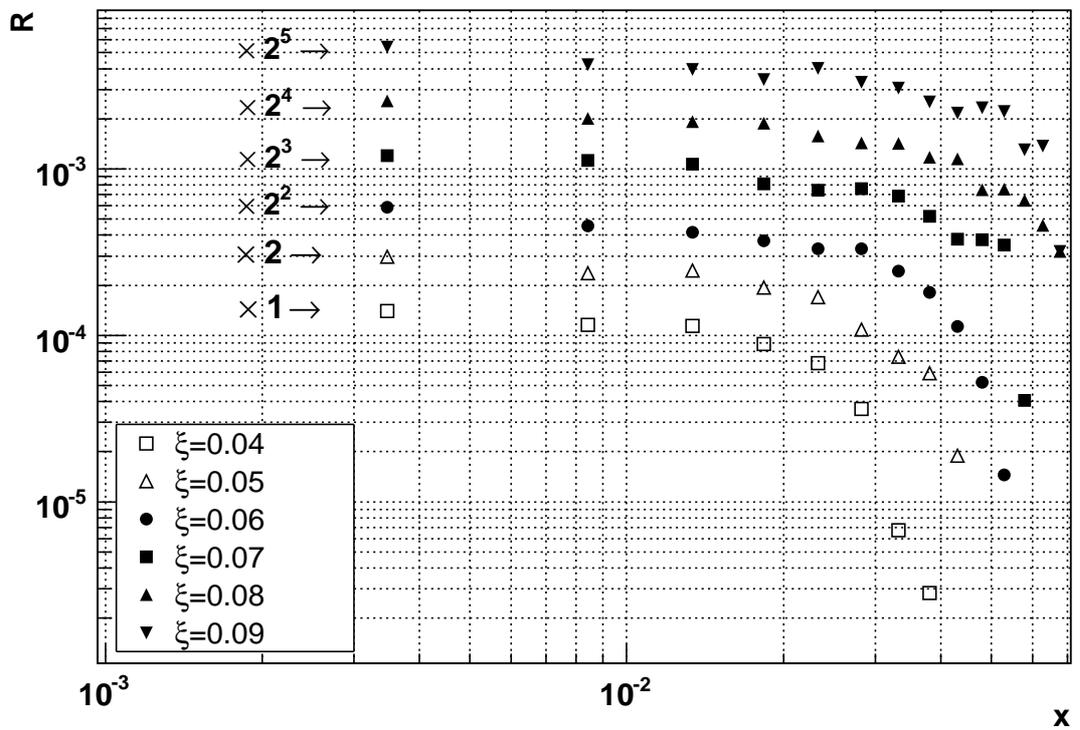}
\vspace*{0pt}
\caption{Ratio of diffractive to non-diffractive dijet event rates as a function of x (momentum
fraction of parton in $p$).}
\label{fig:our}
\end{center}
\end{figure}

\begin{figure}
\begin{center}
\includegraphics[scale=0.6,angle=-90]{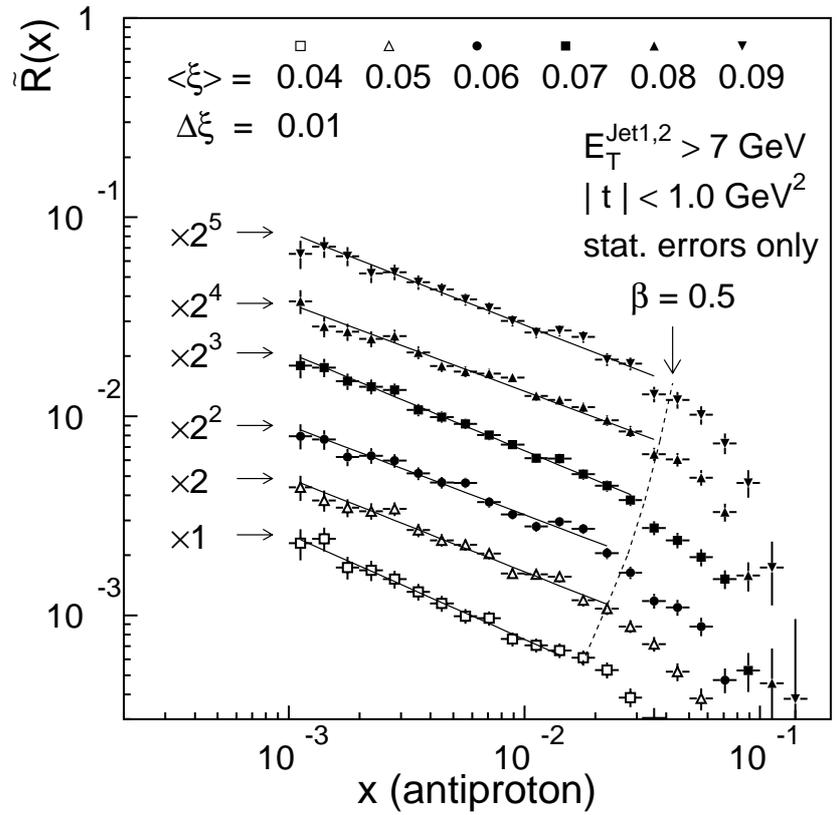}
\vspace*{0pt}
\end{center}
\caption{Ratio of diffractive to non-diffractive dijet event rates as a function of x (momentum
fraction of parton in $p$). The solid lines are fits to the form $R(x) = R_0 (x/0.0065)^{-r}$ for $\beta<0.5$.}
\label{fig:CDF}
\end{figure}

One-to-one interconnection between LCM and SCI models makes SCI
model more predictive, so we can try to describe dependencies of
ratio  of diffractive to non-diffractive events on both Bjorken
and Feynman variables. Results are shown in Fig.\ref{fig:our}, to
be compared with experimental data, which are shown in
Fig.\ref{fig:CDF}, taken from \cite{Affolder:2000vb}.

\newpage

Qualitative agreement is clear, but there are a lot of
quantitative divergences arisen from the oversimplified SCI (GAL)
model assumptions. The direct way to solve this problem is to
enter the correct Couloumb propogator instead of
half-phenomenological probability of soft colour interaction.

\section*{Acknowledgments}
This work was partly supported by RFFI grant 07-07-96410-p. We
thank Prikhodko N.V. for useful discussions.

\end{document}